\documentclass[journal,a4paper,twocolumn,12pt]{IEEEtran}

\usepackage{setspace}
\usepackage{lineno}
\modulolinenumbers[5]
\usepackage{amsthm}
\theoremstyle{definition}
\newtheorem{exmp}{Example}[section]

\newtheorem{definition}{Definition}[section]
\newtheorem{axiom}{Axiom}

\newtheorem{convention}{Convention}[section]

\usepackage[USenglish]{babel} 
\usepackage[T1]{fontenc}
\usepackage{amsfonts}
\usepackage{amssymb}
\usepackage{rotating}
\usepackage{multirow}
\usepackage{booktabs}
\usepackage[]{algorithm2e}
\usepackage{blindtext}
\usepackage{url}
\usepackage{balance}
\usepackage[colorlinks]{hyperref} 
\hypersetup{
 colorlinks=true,
 citecolor=violet,
 linkcolor=blue,
 urlcolor=cyan}

\usepackage{ifluatex}
\ifluatex
\usepackage{fontspec}
\defaultfontfeatures{Ligatures=TeX}
\usepackage[]{unicode-math}
\unimathsetup{math-style=TeX}
\else 
\fi 
\ifluatex\else\usepackage{stmaryrd}\fi

\usepackage{url}
\usepackage{balance}
\usepackage{amsmath}
\usepackage{array}
\usepackage{graphicx}

\usepackage{rotating}
\usepackage{mdwtab}
\usepackage{subfigure}
\usepackage{stfloats}
\usepackage{url}
\usepackage{calc}
\usepackage{curves}
\usepackage{ebezier}
\usepackage{eepic}
\usepackage{epic}
\usepackage{multiply}
\usepackage{pgf}
\usepackage{rotating}
\usepackage{xcolor}
\usepackage{xtab}
\usepackage{datetime}
\usepackage{multirow}
\usepackage[usenames,dvipsnames]{pstricks}
\usepackage{epsfig}
\usepackage{lscape}
\settimeformat{ampmtime}
\usepackage{balance}
\usepackage{mathtools, cuted}

\begin{document}
%
\title{Comprehensive Fuzzy Turing Machines, An Evolution to the Concept of Finite State Machine Control}
%
%
%

\author{Najmeh Ahang,
      Amin Torabi Jahromi,
        and Mansour Doostfatemeh 
\thanks{Amin Torabi Jahromi is with the Department
of Electrical and Computer Engineering, Persian Gulf University, Bushehr, Iran,
 e-mail: a.torabi@pgu.ac.ir}
\thanks{Najmeh Ahang and Mansour Doostfatemeh are with the Department of Mathematics, Shiraz University, Shiraz, Iran e-mail: \{najmeh.ahang,dfatemeh\}@shirazu.ac.ir}
}

\maketitle

\begin{abstract}
 The Turing machine is an abstract concept of a computing device which introduced new models for computation. The idea of Fuzzy algorithms defined by Zadeh and Lee \cite{lee1996note} was followed by introducing  Fuzzy  Turing Machine (FTM)  to create a platform for a new fuzzy computation model  \cite{SANTOS1970326}. Then,  in his investigations on its computational power, Wiedermann showed that FTM is able to solve undecidable problems \cite{wiedermann2004characterizing}.
His suggested FTM structure, which highly resembles the original definition was one of the most well-known classical definitions of FTM lately. 

To improve some of its weaknesses and vague points which will be discussed extensively in this paper,  we will develop a more complete definition for fuzzy Turing machines.  Our proposed definition of FTM, which encompasses the conventional definition, is motivated from the definition of General Fuzzy Automata (GFA) introduced by Doostfatemeh and Kremer \cite{doostfatemeh_new_2005}. As it improved the conventional definition of fuzzy automata, especially the problem of membership assignment and multi-membership resolution, we also improved the same aspects of FTM through the definition of Comprehensive Fuzzy Turing Machine (CFTM). In addition, we address on some possible vaguenesses in FTM was not the subject of focus in fuzzy automata. 
As example, we investigate the issue of multi-path and multi-direction which are possible in case of non-determinism. Finally, we show the simplicity, applicability and computational efficiency  of the CFTM through an explanatory example. 
\end{abstract}


\section{Introduction}\label{Introd}

Incorporation of Fuzzy sets concepts in various branches of science and technology has led to their applicability and flexibility. Although the computational complexity has   increased,  the results has become more accurate and closer to the real world application requirements.  
In computer science, the combination of fuzzy logic and computational systems has resulted to new more effective and complex computational methods. Fuzzy automata  was   the result of incorporation  of fuzzy logic into automata theory. Another computational concept which was introduced and well developed in past decades was Turing Machine (TM) followed by its fuzzy counterpart, Fuzzy Turing Machine (FTM). Alen Turing introduced the concept of TM with the claim that it is as powerful as the human mind. Years later following the introduction of Fuzzy Turing Machine (FTM) and investigation of its computational power, Wiedermann showed that FTM is much more powerful than classical TM, and claimed that FTM has unique capabilities such as modeling and solving \textit{undecidable} problems \cite{wiedermann2004characterizing}. This fact reaffirms the new capabilities of FTM through which many fuzzy algorithms are implementable and many fuzzy languages are accepted. However,  recent investigations   introduced some  languages which was not possible to be accepted by an FTM with its current form of definition \cite{gerla2018fuzzy}. From there, Gerla concludes that conventional FTM is not eligible to be Universal Fuzzy Turing Machine \cite{wiedermann2004characterizing},\cite{gerla2018fuzzy}.

Wiedermann claimed that the conventional fuzzy Turing machine   to be capable of accepting Recursive Enumerable (R.E.) sets and co-R.E. sets \cite{wiedermann2004characterizing,wiedermann_fuzzy_2002}. He also concluded that these machines are able to solve the halting problem.  In \cite{bedregal2008computing}, the  Wiedermann's above statement is investigated by Bedregal and was proved that is  not completely correct. He then  gave a characterization of the class of R.E. sets in terms of associated fuzzy languages accepted by fuzzy Turing machines leading to the nonexistence of a  universal fuzzy Turing machine \cite{farahani2018meta}. 



The rest of the paper is organized as follows;
In the next section, we look into the definition of Wiedermann's Fuzzy Turing Machine (FTM) as the standard classical definition of FTM.  We will also study  its strength and weaknesses. Then, in section. \ref{MMR} we will  develop a more complete formulation for fuzzy Turing machines, cover those vague aspects of the conventional definitions, and propose our own definition of FTM named as Comprehensive Fuzzy Turing Machine (CFTM) which is motivated from the definition of General Fuzzy Automata (GFA). 
In the light of Generalized Fuzzy Automata (GFA) proposed by \cite{doostfatemeh_new_2005}, we developed a more complete definition for two problems already existed in fuzzy Turing machines which covers those faint faces of the membership assignment and multi-membership resolution problem for the states.  As a result, membership values are no longer associated  with IDs and they are directly associated with states. Moreover, for each time step, these membership values are calculated based on the current membership values of states and active transitions and are assigned to the successor states. Due to nondeterminism, there is always a possibility that more than one membership values are assigned to a single state. To resolve the membership assignment problem, we defined a multi-membership resolution function similar to the one existed in the GFA definition.  

However, we noticed that in FTMs, the membership assignment is not the only vague issue. Each active transition requires the machine to move its head in a specific direction and also mandates a predefined symbol to be written on the tape. Therefore, at each time step it is usually more than one symbol to be written on the tape and also more than one direction for the machine to move. 
Hence, in Section .\ref{MMR} we defined two more functions to resolve the above mentioned issues, multi-direction     and multi-symbol resolution functions to decide on  a single direction and 
a single head movement based on the weight of the active transitions and the membership values of their predecessor states. It is easy to prove that each conventional fuzzy Turing machine can be modeled in the form of the novel Comprehensive Fuzzy Turing Machine (CFTM).
Lastly, some comparison on the volume of calculations on conventional FTM and the novel CFTM is performed.

\section{Deficiencies in Conventional FTM Definition}

In conventional definition of FTM, there is a key concept called instantaneous description (ID) of FTM $\textbf{T}$ working on the string $w$ at time $t \geqslant 0$ that it is defined as ``a unique description of the machine's tape content, its state, and the position of the tape head after performing the $t$th move on the input $w$''. 
 Also, there is a function $\mu$ which assigns a weight in $[0,1]$ to each transition $\delta  \in  \Delta$. 
In the 5-tuple $\delta = ({q_1},a,{q_2},b,D)$, $q_1$ and $q_2$ are current state and next state, respectively. The symbol $a$ is the input symbol just read by the head from the tape. The symbol $b$ is the symbol which will be written on the tape by the active transition, and $D$ is the  direction of head movement.
 Then, each ID is assigned a membership value which is calculated based on the transition weight. It means each ID has a membership value calculated from the path it is reached from the previous ID. But, how about the states? Are they assigned any membership values during the process as it was conventionally common in fuzzy automata?
 
  In the following example we followed the calculations of a conventional FTM where the calculations are ID-based for few time steps.
  
 \begin{figure*}[h]
	\centering
  \includegraphics[width=0.95\textwidth]{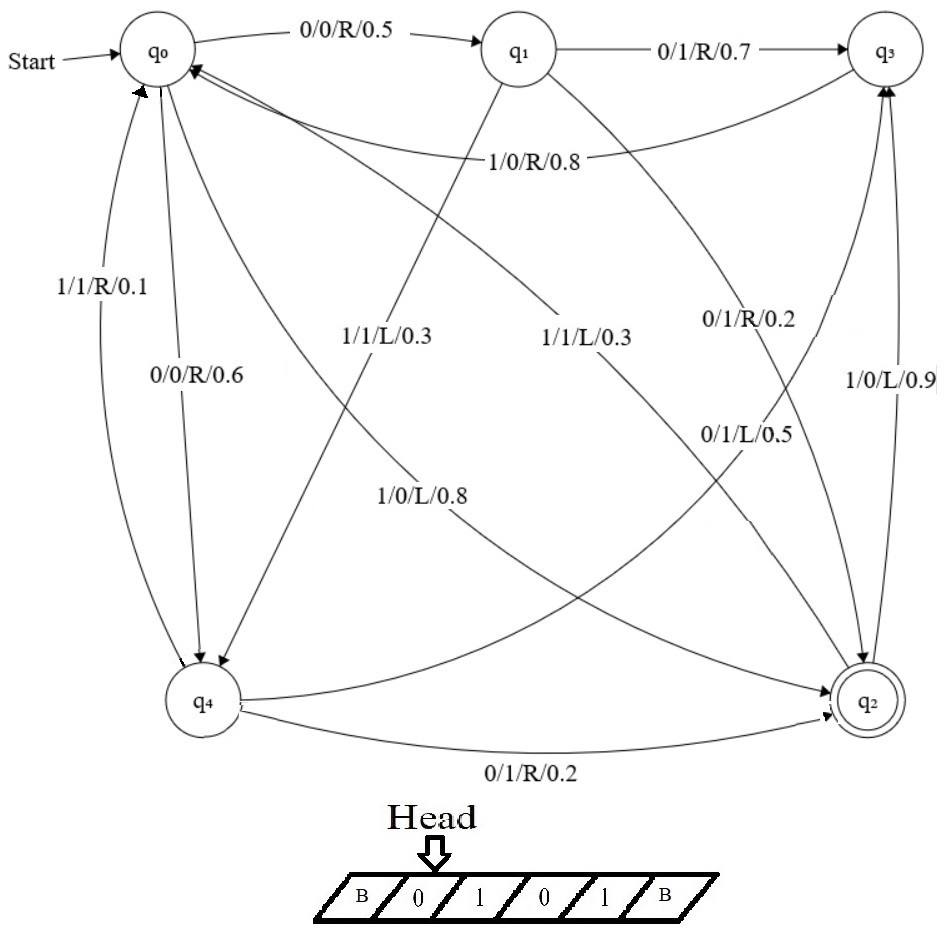}\\
  \caption{A Nondeterministic Fuzzy Turing Machine }
  \label{WiedermannFTM-Example_Lab}
\end{figure*}

Obviously, the amount of calculations are exhaustive and in case of nondeterminism, the possibility of infinite loop is high. 

\begin{exmp}\label{FTMTrace}
In this example we investigate through calculations the amount of  ID-based computations and the complexity of the conventional FTM in \cite{wiedermann2004characterizing} and presented in Fig. \ref{WiedermannFTMtrace_Lab}. Each of the rectangles represent an ID in each time step. Due to huge amount of  calculation, we stopped it after 4 time steps to save space. Yet, it is worth noting that the calculations wont be reaching to their end even after 9 time steps. \\

The original configuration of the tape and  initial head position in Fig. \ref{WiedermannFTM-Example_Lab} is:\\ \begin{tabular}{|c||c||c|c|c|c|}
\hline 
\rule[-2ex]{0pt}{5.5ex}  B & 0 & 1 &0 & 1 & B  \\ 
\hline 
\end{tabular}.\\
The FTM $\textbf{T}$ has the following details:\\
$Q=\{q_0,q_1,q_2,q_3,q_4 \}$, 
$\Sigma=\{0,1\}$, 
$q_0$: start state, and
$q_2$: final state.\\  

\begin{figure*}[th]
	\centering
  \includegraphics[width=.95\textwidth]{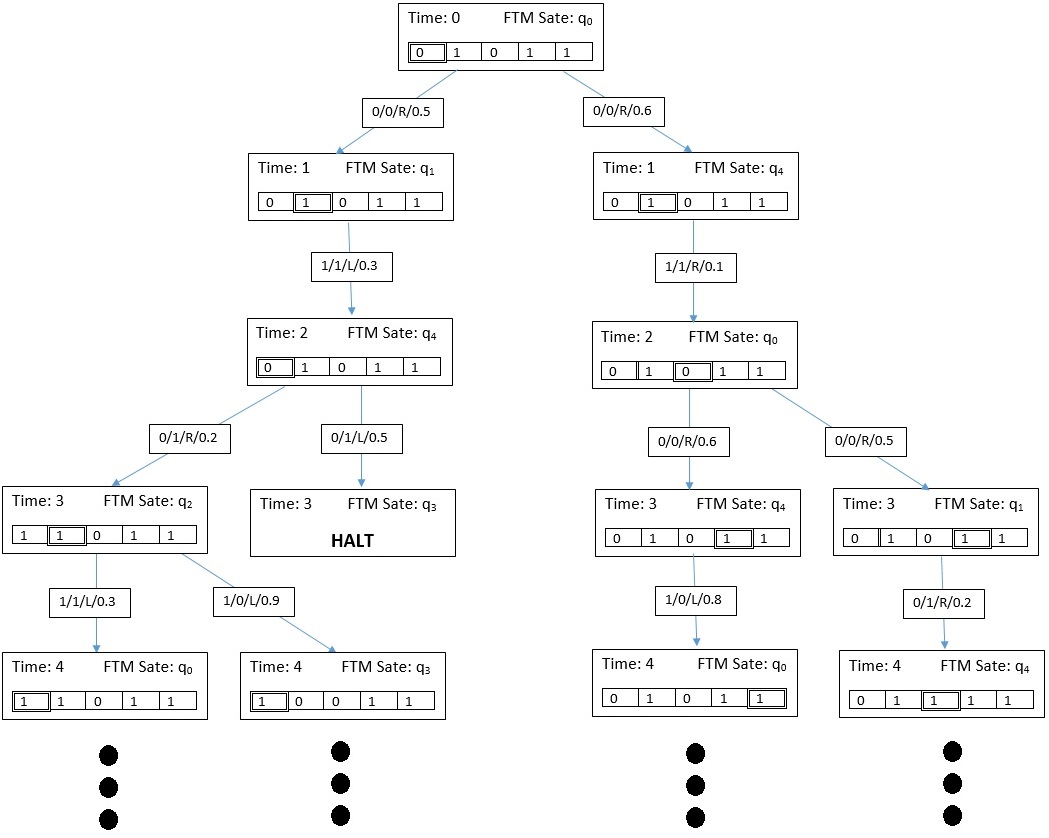}\\
  \caption{A Nondeterministic Fuzzy Turing Machine calculations using conventional FTM ID-based method. Each rectangle represents an ID. }
  \label{WiedermannFTMtrace_Lab}
\end{figure*}

At time step $t=0$, the input symbol is ``$0$'' and the machine starts at state $q_0$. Therefore,
there are two possible moves for the FTM finite state control;
 via transitions $(q_0,0,q_1,0,R,0.5)$, and $(q_0,0,q_4,0,L,0.6)$.
Therefore the next IDs in two branches will be:\\
$Q_1$ in which, machine moves to $q_1$ via a transition with  weight $0.5$ and tape will be: \begin{tabular}{|c|c||c||c|c|c|}
\hline 
\rule[-2ex]{0pt}{5.5ex}  B & 0 & 1 &0 & 1 & B   \\ 
\hline 
\end{tabular} \\
$Q_2$  in which, machine moves to $q_4$ via a transition with weight $0.6$  and tape is \begin{tabular}{|c|c||c||c|c|c|}
\hline 
\rule[-2ex]{0pt}{5.5ex}  B & 0 & 1 &0 & 1 & B  \\ 
\hline 
\end{tabular} \\

The calculations for the next time steps are represented in Fig. \ref{WiedermannFTMtrace_Lab}. It is clear that there are many time steps of calculations required for the conventional FTM. By the time all of the branches reach to \textit{Halt} mode, i.e. there is no more moves possible  or the string reaches to its end, the branches stop growing and the tree would be ready for truth degree calculations. By tracing each individual branches and considering the weight of each transition passed through the branch, the weight for that branch tip final ID would be determined. In case of more than one final ID, the maximum of truth degrees will determine the truth degree associated with that input string.$\square $

\end{exmp}

In conventional definition,
 it is  only the transitions that have weights  and assign membership values to the successor IDs all the way to final ID. The method of computation is very similar to transition-based method introduced in \cite{mordeson2002fuzzy} and well investigated in \cite{doostfatemeh_new_2005}. 
One might address the conventional FTM method for truth degree assignment as ``ID-based method''.  
Although the transition-based method can work well for fuzzy Turing machine realized to accept certain types of fuzzy grammars, it has some disadvantages which makes it unsuitable for many applications as its focus is mostly on the acceptor mode of FTMs. In addition to high computational load in simple FTM, as presented in Example. \ref{FTMTrace}, to study other consequences of transition-based membership, refer to \cite{doostfatemeh_new_2005}.

 Wiedermann definition of FTM operation  resembles the transition-based method where only   transition weights are considered in the assignment of truth degree to the final ID of FTM, and  the \textit{mv}'s of states are not considered and discussed. 

In order to generalize the definition to an applicational one, we follow the methodology in General  Fuzzy Automata (GFA) developed in \cite{doostfatemeh_new_2005} and incorporate a new function in the definition of conventional FTM considering both the transition weight and the \textit{mv} of the predecessor state to assign membership values to the states rather than final IDs. 


Introducing their GFA, ِDoostfatemeh and Kermer devised a method for fuzzy  calculations that moved forward the fuzzy automata calculations to become best suited to practical issues \cite{doostfatemeh_new_2005}.
As seen in natural processes, the phenomenon that occurs at a later time (time $t+1$) is affected by the steps and events that have taken place at the present time (time  $t$).
Therefore, it is reasonably expected that the goal to be achieved in the next period  in a fuzzy automata would be the product of steps taken up to the present time. Precisely, in GFA, the membership value of the next state not only depends on the weight of the active transition, but also incorporates the \textit{mv} of the current state as well. 
Hence, same method might be utilized on the conventional ID-based method in FTMs.

\section{State Membership Assignment in FTMs}\label{FTMMembAssign}

Based  on what we discussed in previous section about GFA, to assign a membership value to a next state, both \textit{mv} of current state  and the weight of the active transition have to be  effective in \textit{mv} calculations.  Hence, we suggest a function which incorporates these two values to assign a membership value to the next state.  There are various options for this function which can be opted based on the application. In the following, we bring some conventions to simplify the presentations. 
\begin{convention}
${\mu}^t(q_m)$ refers to the unique \textit{mv} of the state $q_m$ at time $t$. 
\end{convention} 
\begin{convention}
$\mu _{{q_m}}^t$ refers to the set of \textit{mv}'s associated with the multi-membership state $q_m$ at time $t$. 
\end{convention} 
\begin{convention}
By successor (and predecessor)  state $q_i$,  we mean the states which follow $q_i$ (or are followed by $q_i$) considering a single input symbol  read from the tape at the current time. 
\end{convention}

\begin{convention} In a sample FTM $\textbf{T}$, \hfill
\begin{itemize}
\item $Q:$ Set of states.
\item $\Sigma$ Set of tape symbols.
\item $\Delta$ is the set of all transitions.
\item $\delta: $ is a function with the following definition: $\delta: Q \times \Sigma \times Q \times \Sigma \times \{-1,0,1\} \rightarrow [0,1]$. For example, the weight of the transition ($q_i,a,q_j,b,d$) is $\delta (q_i,a,q_j,b,d)$. $\square$

\end{itemize}
\end{convention}

\begin{figure*}[h]
	\centering
  \includegraphics[width=0.57\textwidth]{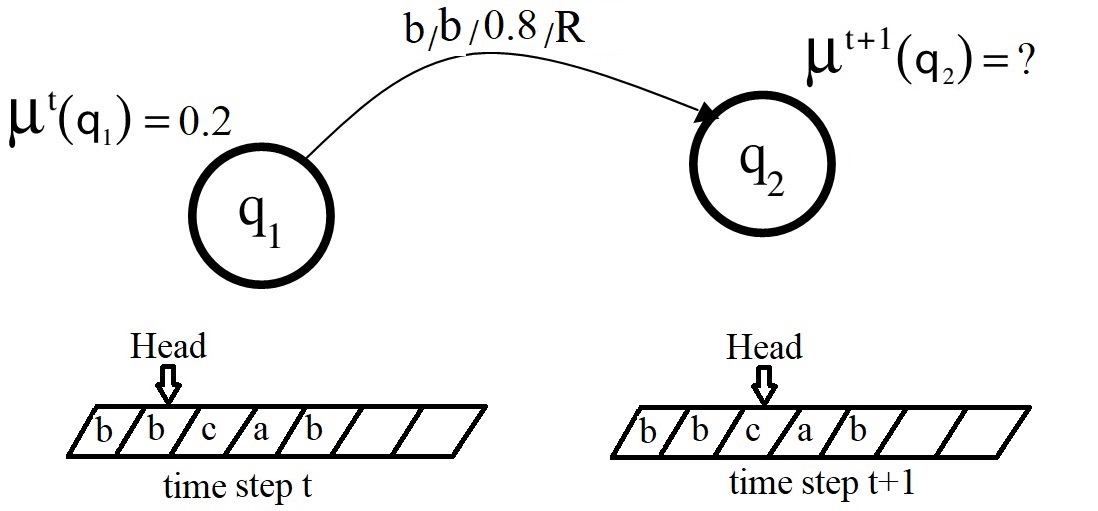}\\
  \caption{An active  transition of a Fuzzy Turing Machine at time step $t$ trying to assign a membership value to  a next state}
  \label{FTM_Again_Lab}
\end{figure*}

Now, we define a new transition function ${\tilde \delta }$, which is called augmented transition function, as follows:
\begin{equation}
\tilde \delta :(Q \times [0,1]) \times \Sigma  \times Q \times \Sigma  \times D\xrightarrow{{{F_1}(\mu ,\delta )}}[0,1]
\end{equation}
$\tilde \delta$ assigns to  the successor state (reached from its predecessor) a value in the  interval [0,1] via function ${F_1}(\mu ,\delta )$ defined as follows.
\begin{definition}
(Membership assignment function) is a mapping function which is applied via \textit{augmented transition function} $\tilde \delta$ to assign \textit{ mv}'s to the active states.
\begin{equation}
{F_1}:[0,1] \times [0,1] \to [0,1]
\end{equation}
Function ${F_1}(\mu ,\delta )$ has two arguments as stated above:
\begin{enumerate}
\item $\mu$: the \textit{mv} of a predecessor;
\item $\delta$: the weight of a transition. 
\end{enumerate}
\begin{equation}
\begin{gathered}
  {\mu ^{t + 1}}({q_j}) = \tilde \delta (({q_i},{\mu ^t}({q_i})),{a_k},{q_j},{b_k},d) \hfill \\
   = {F_1}({\mu ^t}({q_i}),\delta ({q_i},{a_k},{q_j},{b_k},d)) \hfill \\ 
\end{gathered} 
\end{equation}
\end{definition}
which means that the \textit{mv} of the state $q_j$ at time $t+1$ is computed by function $F_1$ using both the \textit{mv} of $q_i$ at time $t$ and the weight of the active transition upon input $a_k$, output $b_k$, and direction $d$.\\
$F_1$ should satisfy the following requirements: 
\begin{axiom}
$0 \leqslant {F_1}(\mu ,\delta ) \leqslant 1$
\end{axiom}
\begin{axiom}
${F_1}(0,0) = 0$ and ${F_1}(1,1) = 1$. 
\end{axiom}
It is clear that $F_1$ function is more flexible and applicational compared to the conventional ID-based method.  It  provides a more suitable platform for  generalization of fuzzy computations in our version of FTM. Refer to \cite{doostfatemeh_new_2005} for more details and discussion on the superiority of $F_1$ definition.

\begin{exmp}
In Fig. \ref{FTM_Again_Lab}, let $F_1(\delta,\mu)= \min(\delta,\mu)$.
As we know, $\mu^t(q_1)=0.2$ and $\delta(q_1,b,q_2,b,R)=0.8$ which yields: 
$\mu^{t+1}(q_2)=\tilde \delta (({q_1},{0.2}),b,{q_2},b,R)={F_1}(\mu^t({q_1}),\delta(q_1,b,{q_2},b,R)=\min(0.2,0.8)=0.2.\square$
\end{exmp}

There are various choices for the function $F_1$. However, the best strategy is always determined by the specific application. In the following, we mention just some  examples as suggested in \cite{doostfatemeh_new_2005}. 

\begin{itemize}
\item ${F_1}(\mu ,\delta ) = Mean(\mu ,\delta ) = \frac{{\mu  + \delta }}{2}$
\item ${F_1}(\mu ,\delta ) = GMean(\mu ,\delta ) = \sqrt {\mu .\delta } $
\item ${F_1}(\mu ,\delta ) = \left\{ {\begin{array}{*{20}{c}}
  {\max (\mu ,\delta )}&{t < {t_i}} \\ 
  {\min (\mu ,\delta )}&{t \geqslant {t_i}} 
\end{array}} \right.$
\item $\begin{array}{*{20}{c}}
  {{F_1}(\mu ,\delta ) = \min \left[ {1,{{({\mu ^\omega } + {\delta ^\omega })}^{1/\omega }}} \right]}&{\omega  > 0} 
\end{array}$ (Yager class of t-conorms \cite{klir1995fuzzy})
\end{itemize}

It is obvious that  ID-based membership assignment  to the next configuration can be considered as a special case  where  $F_1(\mu,\delta)=\delta$. This fact, enables our version of FTM to encompass the conventional versions of FTM. 

\begin{exmp}
Let us familiarize ourselves with the  FTM fuzzy calculations. In this example, the deterministic FTM includes $Q = \{ {q_0},{q_1},{q_2},{q_3},{q_4},{q_5}\} $,  $\Sigma  = \{ a,b,c\}$, $\Gamma  = \{ a,b,c,B\}$, $q_0=$ start state, and  $F = \{ {q_5}\}$.
In this example, to do the states' membership value calculations in  Fig.  \ref{F1_Example_Lab} FTM, we apply $F_1(\mu,\delta)= \frac{{\mu  + \delta }}{2}$. 

\begin{figure*}[h]
	\centering
  \includegraphics[width=0.77\textwidth]{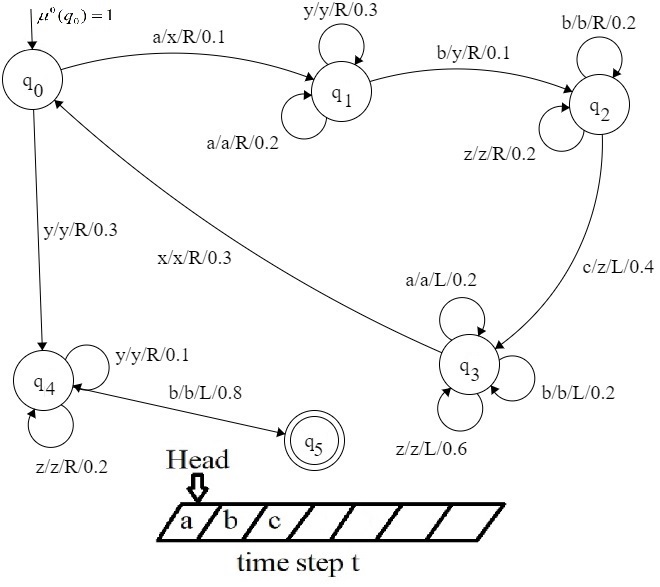}\\
  \caption{A deterministic FTM which accepts the language $L=\{a^nb^nc^n\}$ for $n \geqslant 1 $}
  \label{F1_Example_Lab}
\end{figure*}

The following table carries the simulation results for a glance.

\begin{table*}[htbp]
  \centering
  \caption{The FTM in Fig. \ref{F1_Example_Lab} working on the string $abc$}
\begin{tabular}{||c|c|c|c|c||}
\hline 
time & 0 &  1 & 2 & 3 \\ 
\hline 
input & $\epsilon$ & a & b & c \\ 
\hline 
$Q_{act}$ & $q_0$ & $q_1$ & $q_2$ & $q_3$ \\ 
\hline 
\textit{mv} & 1 & 0.55 & 0.325 & 0.3625 \\ 
\hline 
Symbol to Write (output) & - & x & y & z \\ 
\hline 
Direction & - & R & R & L \\ 
\hline 
\end{tabular} 
  \label{tab:addlabejghjhl}%
\end{table*}%

The performed calculations to fill the above table are as follows:

At time step $t=0$, $input=\epsilon$ (empty input),  $\mu^{t_0}(q_0)=1$.\\
At time step $t=1$, $input=$a, \begin{tabular}{|c||c||c|c|c|}
\hline 
\rule[-2ex]{0pt}{5.5ex}  B & a & b & c &  B  \\ 
\hline 
\end{tabular} 
\[\begin{gathered}
  {\mu ^{{t_1}}}({q_1}) = {F_1}({\mu ^{{t_0}}}({q_0}),\delta ({q_0},a,{q_1},x,R)) \hfill \\
   = {F_1}(1,0.1) = \frac{{1 + 0.1}}{2} = 0.55] \hfill \\ 
\end{gathered} \]
At time step $t=2$, $input=$b, \begin{tabular}{|c|c||c||c|c|}
\hline 
\rule[-2ex]{0pt}{5.5ex}  B & x & b & c &  B  \\ 
\hline 
\end{tabular}\[\begin{gathered}
  [{\mu ^{{t_2}}}({q_2}) = {F_1}({\mu ^{{t_1}}}({q_1}),\delta ({q_1},b,{q_2},y,R)) \hfill \\
   = {F_1}(0.55,0.1) = \frac{{0.55 + 0.1}}{2} = 0.325] \hfill \\ 
\end{gathered} \]
At time $step t=3$, $input=$c, \begin{tabular}{|c|c|c||c||c|}
\hline 
\rule[-2ex]{0pt}{5.5ex}  B & x & y & c &  B  \\ 
\hline 
\end{tabular}  \[\begin{gathered}
  {\mu ^{{t_3}}}({q_3}) = {F_1}({\mu ^{{t_2}}}({q_2}),\delta ({q_2},c,{q_3},z,L)) \hfill \\
   = {F_1}(0.325,0.4) = \frac{{0.325 + 0.4}}{2} = 0.3625] \hfill \\ 
\end{gathered} \]
\end{exmp}

\section{Multi-Membership, Multi-Symbol, and Multi-Direction  Resolution}\label{MMR}

One of the interesting issues which occurs in nondeterministic FTM, similar to its ancestor FFA, is simultaneous  transitions to the same state. In previous section, we addressed the membership assignment problem, defining the $F_1$ function which incorporates \textit{mv} of  predecessor state and transition weight to calculate the membership value of the next state. Because of nondeterminism, in some cases we have several membership values to be assigned to a successor state when there are  several simultaneous transitions  to that state. The question is what will be the actual membership value of the next state?

\begin{figure*}[h]
	\centering
  \includegraphics[width=0.4\textwidth]{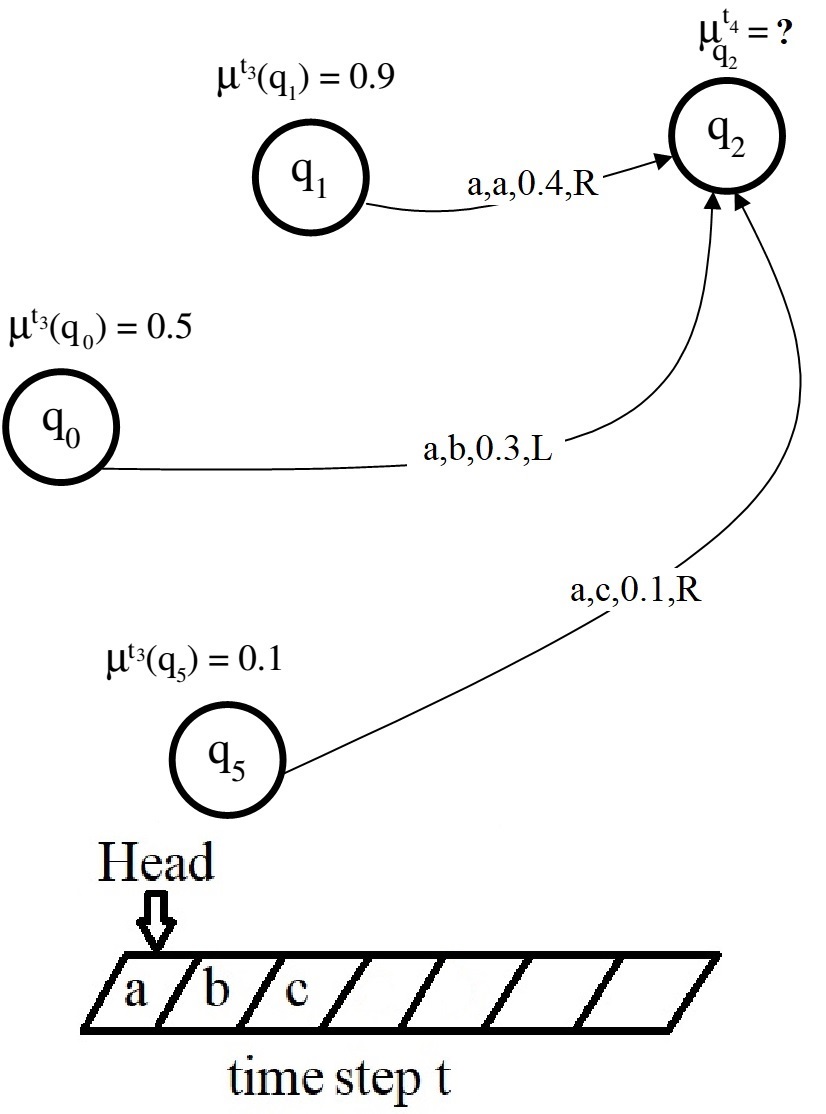}\\
  \caption{A part of a nondeterministic fuzzy Turing machine which depicts multi-membership}
  \label{MultiMembershipProblem_Lab}
\end{figure*}

For example, in Fig. \ref{MultiMembershipProblem_Lab}, all states have \textit{mv}'s and all transitions have weights. If we consider $F_1(\mu,\delta)= GMean(\mu,\delta)= \sqrt {\mu .\delta }$, then: 

At $t=t_3$, $input=$a, and original state $q_1$: \[\begin{gathered}
  [{\mu ^{{t_4}}}({q_2}) = {F_1}({\mu ^{{t_3}}}({q_1}),\delta ({q_1},a,{q_2},a,R)) \hfill \\
   = {F_1}(0.9,0.4) = \sqrt {0.9 \times 0.4}  = 0.6] \hfill \\ 
\end{gathered} \]

Again, at $t=t_3$, $input=$a, and original state $q_0$: \[\begin{gathered}
  {\mu ^{{t_4}}}({q_2}) = {F_1}({\mu ^{{t_3}}}({q_0}),\delta ({q_0},a,{q_2},b,L)) \hfill \\
   = {F_1}(0.5,0.3) = \sqrt {0.5 \times 0.3}  = 0.387] \hfill \\ 
\end{gathered} \]
And again, at $t=t_3$, $input=$a, and original state $q_5$: \[\begin{gathered}
  {\mu ^{{t_4}}}({q_2}) = {F_1}({\mu ^{{t_3}}}({q_5}),\delta ({q_5},a,{q_2},c,R)) \hfill \\
   = {F_1}(0.1,0.1) = \sqrt {0.5 \times 0.3}  = 0.1] \hfill \\ 
\end{gathered} \]

Therefore, $q_2$ gets activated at $t_4$ from three different paths with three different \textit{mv}'s $\{ 0.6, 0.387, 0.1 \}$, while only a single \textit{mv} has to be assigned to $q_2$. This issue is called multi-membership problem.

To the best of our knowledge, 
available literature and research  have no solution  to characterize the operation of the FTM when it comes to multi-membership problem.   Fortunately, there are methodologies in \cite{doostfatemeh_new_2005} for fuzzy automata to calculate the  membership value of the states at time $t+1$ even in the existence of multi-membership value problem. The idea can be extended to FTM with some minor changes.

Motivated by the method presented in \cite{doostfatemeh_new_2005}, we  define some conventions to provide a suitable platform to resolve the multi-membership  problem.

In conventional FTM,  transition $(q_i, a_k, q_j, b_k, d)$, includes $q_i$ which represents the current state, $a_k$ that is the incoming symbol (the symbol which is present on the current position of the tape head), $q_j$ which is the next state, $b_k$ is the output symbol which is going to be written on the tape, and $d$ is the direction of   the head movement. In our  proposed version of the FTM we utilize the following conventions:

\begin{convention}
Set of all transitions of  fuzzy Turing machine $\textbf{F}$ is denoted by $\Delta_{\textbf{F}}$.
\end{convention}

\begin{definition}
(Successor set): $Q_{Succ}(q_i,a_k)$ is the set of all destination states such as $q_j$ in all  transitions with origin  $q_i$ like $(q_i,a_k,q_j,b_k,d)$ when the input symbol is $a_k$. 
\begin{equation}
\begin{gathered}
  {Q_{Succ}}({q_i},{a_k}) = \{ {q_j}|({q_i},{a_k},{q_j},{b_k},d) \hfill \\
   \in {\Delta _{\mathbf{F}}}{\text{ when the  input symbol is }}{a_k}\}  \hfill \\ 
\end{gathered} \end{equation}
\end{definition}

\begin{definition}(Predecessor set):  $Q_{Pred}(q_j,a_k)$ is the set of all states followed by $q_j$ following the input symbol $a_k$.
\begin{equation}
\begin{gathered}
  {Q_{Pred}}({q_j},{a_k}) = \{ {q_i}|({q_i},{a_k},{q_j},{b_k},d) \hfill \\
   \in {\Delta _{\mathbf{F}}}{\text{ when the input symbol is }}{a_k}\}  \hfill \\ 
\end{gathered} 
\end{equation}
\end{definition}

\begin{definition}
(Active state set) After entering input $a_k$ at time $t$ to the FTM, there are some states that have at least one transition directed to them on input symbol $a_k$. The  set of these states along with their membership values is called \textit{active state  set at time $t$} which is  denoted as $Q_{Act}(t)$. Note that $Q_{Act}(t)$ is a fuzzy set. 
\end{definition}

\begin{exmp}
In Fig. \ref{MultiMembershipProblem_Lab} after input `$a$' at time $t_3$,  $Q_{Act}(t_4)$  can be calculated as  $\{(q_1,0.9),(q_0,0.5),(q_5,0.1)\}$  which presents clearly a multi-membership problem. $\square$
\end{exmp}

In FTM, overlapping of transitions to state is more problematic than fuzzy automata, since it not only makes the assignment of \textit{mv} to that state ambiguous, but also creates ambiguity to the decision on the direction of head movement
 and the symbol to be written on the tape and they  have to be uniquely determined in a reasonable way.
 
 Referring again to  Fig. \ref{MultiMembershipProblem_Lab}, we notice that in addition to  the multi-membership value problem, we have multi-symbol and multi-direction  problem to be resolved too. 
 As an example, all three active transitions after incoming symbol `$a$', each tries to write its  own suggested symbol  on the tape. Hence, the problem arises that which member of the set $\{a,b,c\}$ should be written on the tape? Similarly, the movement direction of the head suggested by two of the three transitions is \textit{Right} while the other tries to move the head to the \textit{Left}. Again, it will require a proper judgment to be imposed to resolve the multi-direction issue. 
 To the best of our knowledge, these above mentioned issues have never been addressed so far in literature among several available definitions.

  To resolve the  multi-membership, multi-symbol, and multi-direction problem, we  evaluated three options for resolution methods. 
  
\textbf{1)} The first resolution method is based on the  conventional definition for FTM, where transition weights  are involved to assign membership value to the successor IDs -very similar to transition-based membership assignment method. The main concern in this method is the final ``accepting'' ID membership value and not only other ID's or states. Therefore, the path to the accepting  ID is considered to evaluate the degree of acceptance, and a final decision is made  in cases there are more than one path to the accepting ID. Hence, the multi-direction issues were never faced as there is only  one possible path considered  and  a tree is formed from the machine possible movements, refer to example \ref{WiedermannFTMtrace_Lab}. 
 The same problem holds for the symbol   to be  written on the tape, (\textit{multi-symbol problem}). Another aspect of the conventional FTM calculation is the volume of calculation needed to trace each possible path from the initial  to the final ID. Due to possible nondeterminism, at each branch in the automata (at least two active transitions from one  state), another new truth degree calculation branch is initiated and its respective truth degree is considered as a possible candidate for the final truth degree of the input string. 
In this method,  the truth degree assignment is performed only after  each and every new path is finalized. In some cases, it takes many or even infinite calculations for a  simple FTM to determine a truth degree for a string.

\textbf{2) }Core idea of the second resolution method for aforementioned issues is extracted from ambiguity removal idea  discussed in \cite{giles1999equivalence} by Omlin. In his suggested method, when an overlapping problem is observed for a state, a new state is generated for each of the conflicting transitions, and this process is continued until there will be no two conflicting  transitions directed to one single state. In practice, this resolution method causes two major problems:
\begin{itemize}
\item Generation of many new states that change the original finite  control (FC) to a much more complex one. The new FC is no longer identical to the simple initial one and the original form  cannot be distinguished among the numerous newly defined states. This issue is addressed well in \cite{doostfatemeh_new_2005}.
\item Due to considerable number of new states created by this  method, it increases considerably the volume of fuzzy computations, which may lead to impracticality for large fuzzy Turing machines.
\end{itemize}
It is quite obvious that, following the above idea for FTM, a set of new tapes have to be created once a multi-symbol problem is faced.  There are several issues with this method as described below:

\begin{itemize}
\item Each new tape has to be identical to original tape, but they will differ at the place that the head points to at time $t$. From that moment on, since FTM possess new tapes to handle, the transitions of its FC have to be modified accordingly. For example, suppose there are $k$ number of tapes available at time $t$, which mandates the FTM transitions to have $k$ input symbols. Also, suppose for a multi-symbol problem before time step $t$,  $i$ new  tapes are generated. For the new FTM to manage these tapes, each transition needs to have $k+i$ input symbols. This implies that the FC  of the FTM  have to be thoroughly modified, which considerably complicates the calculations and FTM management problem.
 To illustrate the issue, for any instantaneous description (ID) containing a nondeterminism, one might generate a new tape so that the symbol suggested by each active transition be written on the respective new tape and the tape  moves along the suggested direction for the next step. As the number of multi-symbol and multi-direction issues in FTM computations increases, it leads to numerous new tapes which may again lead to FTM blow up. Therefore, generating these tapes, managing their computations, dealing with ever-changing FC are the consequences of this  solution which makes it almost impractical to  implement and compute.
\item In practice, having many new tapes generated with identical content, and moving their heads to a specified location is troublesome. 
\item From that moment onward, each tape will follow its own direction based on  active transition. It simply manifolds the complexity of the multi-symbol problem. 
\end{itemize}

\textbf{ 3)} In the third method, which is our novel approach, we consider a set of active transition(s) in  fuzzy Turing machine at each time step. These 5-tuple transitions $(q_i,a_k,q_j,b_k,d)$ are composed of three parts; current and next state,  symbol read and to be written on the tape, and head movement direction. Suppose that in the above mentioned set,  there are more than one active transitions directed to a next state $q_j$, each requires to: \\1- assign a membership value,\\ 2- determine the direction of head movement, and \\3- write its own symbol on the tape.     
\subsection{\textbf{Multi-membership Resolution}}

We suggest a solution to first problem using another function that we call  $F_2$ or (\textit{multi-membership resolution function}):

\begin{definition}
In FTM, the multi-membership resolution function is a function which combines  \textit{mv}'s of an active state, and produces a unique membership value for a state to be used in the next time step.
$F_2:[0,1]^* \rightarrow [0,1]$.
\end{definition}

Similar to what we suggested for  $F_1$, there are some requirements that $F_2$ has to meet:

\begin{axiom} 

$\begin{gathered}
  0\mathop {{F_2}}\limits_{i = 1}^n ({\nu _i})1 \hfill \\
  {\nu _i} = {F_1}({\mu ^t}({q_i}),\delta ({q_i},{a_k},{q_j},{b_k},d)) \hfill \\ 
\end{gathered} $.
\end{axiom}

\begin{axiom} 
$F2(\emptyset ) = 0$.
\end{axiom}

\begin{axiom} 
$\mathop {{F_2}}\limits_{i = 1}^n ({\nu _i}) = a,$ if $\forall i,{\nu _i} = a$.
\end{axiom}

There can be several options for $F_2$, where the best choice have to be determined by the application under consideration. Some possible candidates are as follows \cite{doostfatemeh_new_2005}:
\begin{itemize}
\item Maximum multi-membership resolution:
\begin{equation}
\begin{gathered}
  {\mu ^{t + 1}}({q_j}) = \mathop {\max }\limits_{i = 1}^n \left[ {\tilde \delta \left( {({q_i},{\mu ^t}({q_i})),{a_k},{q_j},{b_k},d} \right)} \right] \hfill \\
   = \mathop {\max }\limits_{i = 1}^n \left[ {{F_1}\left( {{\mu ^t}({q_i}),\delta ({q_i},{a_k},{q_j},{b_k},d)} \right)} \right] \hfill \\ 
\end{gathered} 
 \end{equation}
\item Arithmetic mean multi-membership resolution: 
\begin{equation}
\begin{gathered}
  {\mu ^{t + 1}}({q_j}) = \left[ {\sum\limits_{i = 1}^n {\tilde \delta \left( {({q_i},{\mu ^t}({q_i})),{a_k},{q_j},{b_k},d} \right)} } \right]/n \hfill \\
   = \left[ {\sum\limits_{i = 1}^n {{F_1}\left( {{\mu ^t}({q_i}),\delta ({q_i},{a_k},{q_j},{b_k},d)} \right)} } \right]/n \hfill \\ 
\end{gathered} 
 \end{equation}
 
\item Geometric mean multi-membership resolution:
\begin{equation}
\begin{gathered}
  {\mu ^{t + 1}}({q_j}) = \sqrt[n]{{\prod\limits_{i = 1}^n {\tilde \delta \left( {({q_i},{\mu ^t}({q_i})),{a_k},{q_j},{b_k},d} \right)} }} \hfill \\
   = \sqrt[n]{{\prod\limits_{i = 1}^n {{F_1}\left( {{\mu ^t}({q_i}),\delta ({q_i},{a_k},{q_j},{b_k},d)} \right)} }} \hfill \\ 
\end{gathered} 
 \end{equation}

\end{itemize}
where $n$ is the number of simultaneous transitions from $q_i$'s to $q_m$ at time $t+1$, and ${q_i} \in {Q_{pred}}({q_m},{a_k})$.

\begin{exmp}
For the membership value calculations of  Fig. \ref{MultiMembershipProblem_Lab}, the results are gathered in a set like  $\{ 0.6, 0.387, 0.1 \}$ which illustrates a simple case of multi-membership problem. In order to resolve   this issue, one  can utilize an  $F_2$ function like Arithmetic mean. Therefore, the actual membership value which will be  assigned to  state $q_2$ is calculated as:
 $\mu^{t+1}(q_2)=({0.6 + 0.387 + 0.1})/3=0.362$
\end{exmp}

\subsection{\textbf{Multi-symbol Resolution}}

To resolve the multi-symbol and multi-direction problem we have to consider some new conventions:

\begin{convention}
Suppose  $a_k$ is an input tape symbol at time $t$. The \textit{active transitions} are those  with the form $({q_i},{a_k},{q_j},{b_k},d)$ whose $\mu^t(q_i)$'s are nonzero. 
\end{convention}

\begin{definition} ( Set of pairs including current active transitions and their weights)
\begin{equation}
\begin{gathered}
  \Delta _{Act}^t({a_k}) = \{ [({q_i},{a_k},{q_j},{b_k},d),{F_1}({\mu ^t}({q_i}), \\ \delta ({q_i},{a_k},{q_j},{b_k},d))]{\text{ }}|({q_i},{a_k},{q_j},{b_k},d) \hfill \\
   \in {\Delta _{FTM}},{\mu ^t}({q_i}) \ne 0, \\ {\text{ and current input symbol from the tape is }}{a_k}\}  \hfill \\ 
\end{gathered} 
\end{equation}

i.e. the set of all active transitions at time step $t$ with regards to the input $a_k$.

\end{definition}

\begin{exmp}\label{DeltaActExmp}
In the FTM depicted in Fig. \ref{MultiMembershipProblem_Lab}, suppose that $F_1$ function be the algebraic product t-norm. The  $\Delta _{Act}^t(a)$ set  will simply be:

\[\begin{gathered}
  \Delta _{Act}^t(a) = \left\{ \begin{gathered}
  \left[ {(a,a,0.4,R),{F_1}(0.9,0.4)} \right], \hfill \\
  \left[ {(a,b,0.3,L),{F_1}(0.5,0.3)} \right], \hfill \\
  \left[ {(a,c,0.1,R),{F_1}(0.1,0.1)} \right] \hfill \\ 
\end{gathered}  \right\} \hfill \\
  {\text{              }} = \left\{ \begin{gathered}
  \left[ {(a,a,0.4,R),(0.9 \times 0.4)} \right], \hfill \\
  \left[ {(a,b,0.3,L),(0.5 \times 0.3)} \right], \hfill \\
  \left[ {(a,c,0.1,R),(0.1 \times 0.1)} \right] \hfill \\ 
\end{gathered}  \right\} \hfill \\ 
\end{gathered} \]
`
\end{exmp}

Each of the active transitions which are members of $\Delta^t_{Act}(a_k)$ suggests a symbol to be written on the tape.
To resolve any confusion about these symbols and to agree upon a  single symbol which will be written on the tape, we define a function $F_3$ as following:

\begin{definition} (Multi-symbol resolution function)
\begin{equation}
F_3:\Delta^t_{Act}(a_k)  \rightarrow \Sigma 
\end{equation}
As is clear, set of pairs of active transitions  $(q_i,a_k,q_j,b_k,d)$,  their  weights $\delta(q_i,a_k,q_j,b_k,d)$, and the membership value of their origin $\mu^t(q_i)$  at time $t$ when the input symbol $a_k$ is read from the tape are required for calculations of the symbol to be written on the tape. The transition details are required because they include the symbol  to be written on the tape and the $\delta$ and $\mu^t(q_i)$  are needed by $F_3$ to determine the  strength of that transition. For the sake of simplicity, let us limit the criterion for choosing the symbol and the direction to be only based on $F_1$
of each transition. But, the method is open for further modifications in cases when $F_3$ needs to be independent of $F_1$.  
\end{definition}

%

There can be several options for $F_3$, where the best choice have to be determined by the application. Some possible candidates might be as follows:
\begin{itemize}
\item The symbol in the active transition  with maximum weight represented in Eq.\ref{F3Formula1}. \\

\begin{strip}

\begin{equation}
 {F_3}\left( {\Delta _{Act}^t({a_k})} \right) = \left\{ \begin{gathered}
  {b_k}|\left[ {\left( {{q_i},{a_k},{q_j},{b_k},d} \right),{F_1}\left( {{\mu ^t}({q_i}),\delta ({q_i},{a_k},{q_j},{b_k},d)} \right)} \right] \in \Delta _{Act}^t({a_k}), \hfill \\
  \forall \left[ {\left( {q{'_i},{a_k},q{'_j},b{'_k},d'} \right),{F_1}\left( {{\mu ^t}({q_i}),\delta (q{'_i},{a_k},q{'_j},b{'_k},d')} \right)} \right] \in \Delta _{Act}^t({a_k}), \hfill \\
  {F_1}\left( {{\mu ^t}({q_i}),\delta (q{'_i},{a_k},q{'_j},b{'_k},d')} \right) < {F_1}\left( {{\mu ^t}({q_i}),\delta ({q_i},{a_k},{q_j},{b_k},d)} \right) \hfill \\ 
\end{gathered}  \right\}
\label{F3Formula1}
\end{equation}

\end{strip}

In case of equal weights, one might select the suggestion of the transition with maximum  membership value of its respective predecesor. 

\item The symbol in the active transition set with maximum scalar cardinality - sigma-count - of membership values of transitions (summation of the weight of transitions that suggest that specific symbol) represented in Eq.\ref{F3Formula2}.\\

\begin{strip}

\begin{equation}
{F_3}\left( {\Delta _{Act}^t({a_k})} \right) = \left\{ \begin{gathered}
  {b_k}|\left[ {\left( {{q_i},{a_k},{q_j},{b_k},d} \right),{F_1}\left( {{\mu ^t}({q_i}),\delta ({q_i},{a_k},{q_j},{b_k},d)} \right)} \right] \in \Delta _{Act}^t({a_k}), \hfill \\
  \forall \left[ {\left( {q{'_i},{a_k},q{'_j},b{'_k},d'} \right),{F_1}\left( {{\mu ^t}({q_i}),\delta (q{'_i},{a_k},q{'_j},b{'_k},d')} \right)} \right] \in \Delta _{Act}^t({a_k}), \hfill \\
  \sum {{F_1}\left( {{\mu ^t}({q'_i}),\delta (q{'_i},{a_k},q{'_j},b{'_k},d')} \right)}  < \sum {{F_1}\left( {{\mu ^t}({q_i}),\delta ({q_i},{a_k},{q_j},{b_k},d)} \right)}  \hfill \\ 
\end{gathered}  \right\}
\label{F3Formula2}
\end{equation}
\end{strip}

\item The symbol in the active transition set with maximum cardinal (number of the transitions that suggest the specific symbol) represented in Eq.\ref{F3Formula3}.\\

\begin{strip}
\begin{equation}
{F_3}\left( {\Delta _{Act}^t({a_k})} \right) = \left\{ \begin{gathered}
  {b_k}|\left[ {\left( {{q_i},{a_k},{q_j},{b_k},d} \right),{F_1}\left( {{\mu ^t}({q_i}),\delta ({q_i},{a_k},{q_j},{b_k},d)} \right)} \right] \in \Delta _{Act}^t({a_k}), \hfill \\
  \forall \left[ {\left( {q{'_i},{a_k},q{'_j},b{'_k},d'} \right),{F_1}\left( {{\mu ^t}({q_i}),\delta (q{'_i},{a_k},q{'_j},b{'_k},d')} \right)} \right] \in \Delta _{Act}^t({a_k}), \hfill \\
  \sum {\left\lceil {{F_1}\left( {{\mu ^t}(q{'_i}),\delta (q{'_i},{a_k},q{'_j},b{'_k},d')} \right)} \right\rceil }  < \sum {\left\lceil {{F_1}\left( {{\mu ^t}({q_i}),\delta ({q_i},{a_k},{q_j},{b_k},d)} \right)} \right\rceil }  \hfill \\ 
\end{gathered}  \right\}
\label{F3Formula3}
\end{equation}
\end{strip}

\end{itemize}

\subsection{\textbf{Multi-direction Resolution}}
Each of the active transitions which are members of $\Delta^t_{Act}(a_k)$ suggests a symbol to be written on the tape as well as requiring the tape head to move along an specific direction.
To resolve any confusion about the head's direction and to decide on one single direction among many,   we define a function $F_4$ as following:

\begin{definition} (multi-direction resolution function)
\begin{equation}
F_4: \Delta^t_{Act}(a_k)  \rightarrow  D
\end{equation}
As is clear from above definition, set of pairs of  active transitions $(q_i,a_k,q_j,b_k,d)$,  their  weights $\delta(q_i,a_k,q_j,b_k,d)$, and the membership value of their origin $\mu^t(q_i)$  at time $t$ when the input symbol $a_k$ is read from the tape ($\Delta^t_{Act}(a_k)$) are incorporated by $F_4$ to choose a direction for head movement.  The rest of the details are identical to $F_3$. 
\end{definition}

There can be several options for $F_4$, where the best choice have to be determined by the application. Some possible candidates are as follows:
\begin{itemize}
\item The direction in the active transition  with maximum weight represented in Eq.\ref{F4Formula1}.\\

\begin{strip}
\begin{equation}
{F_4}\left( {\Delta _{Act}^t({a_k})} \right) = \left\{ \begin{gathered}
  d|\left[ {\left( {{q_i},{a_k},{q_j},{b_k},d} \right),{F_1}\left( {{\mu ^t}({q_i}),\delta ({q_i},{a_k},{q_j},{b_k},d)} \right)} \right] \in \Delta _{Act}^t({a_k}), \hfill \\
  \forall \left[ {\left( {q{'_i},{a_k},q{'_j},b{'_k},d'} \right),{F_1}\left( {{\mu ^t}(q{'_i}),\delta (q{'_i},{a_k},q{'_j},b{'_k},d')} \right)} \right] \in \Delta _{Act}^t({a_k}), \hfill \\
  {F_1}\left( {{\mu ^t}(q{'_i}),\delta (q{'_i},{a_k},q{'_j},b{'_k},d')} \right) < {F_1}\left( {{\mu ^t}({q_i}),\delta ({q_i},{a_k},{q_j},{b_k},d)} \right) \hfill \\ 
\end{gathered}  \right\}
\label{F4Formula1}
\end{equation}
\end{strip}

In case of equal weights, one might select the suggestion of the transition with maximum weight  of its respective predecesor.

\item The direction in the active transition set with maximum scalar cardinality  - sigma-count - of membership values of transitions (summation of the weight of transitions that suggest that specific direction) represented in Eq.\ref{F4Formula2}.\\

\begin{strip}
\begin{equation}
{F_4}\left( {\Delta _{Act}^t({a_k})} \right) = \left\{ \begin{gathered}
  d|\left[ {\left( {{q_i},{a_k},{q_j},{b_k},d} \right),{F_1}\left( {{\mu ^t}({q_i}),\delta ({q_i},{a_k},{q_j},{b_k},d)} \right)} \right] \in \Delta _{Act}^t({a_k}), \hfill \\
  \forall \left[ {\left( {q{'_i},{a_k},q{'_j},b{'_k},d'} \right),{F_1}\left( {{\mu ^t}(q{'_i}),\delta (q{'_i},{a_k},q{'_j},b{'_k},d')} \right)} \right] \in \Delta _{Act}^t({a_k}), \hfill \\
  \sum {{F_1}\left( {{\mu ^t}(q{'_i}),\delta (q{'_i},{a_k},q{'_j},b{'_k},d')} \right)}  < \sum {{F_1}\left( {{\mu ^t}({q_i}),\delta ({q_i},{a_k},{q_j},{b_k},d)} \right)}  \hfill \\ 
\end{gathered}  \right\}
\label{F4Formula2}
\end{equation}
\end{strip}

\item The direction in the active transition set with maximum cardinal (number of the transitions that suggest the specific direction) represented in Eq.\ref{F4Formula3}. \\

\begin{strip}
\begin{equation}
{F_4}\left( {\Delta _{Act}^t({a_k})} \right) = \left\{ \begin{gathered}
  d|\left[ {\left( {{q_i},{a_k},{q_j},{b_k},d} \right),{F_1}\left( {{\mu ^t}({q_i}),\delta ({q_i},{a_k},{q_j},{b_k},d)} \right)} \right] \in \Delta _{Act}^t({a_k}), \hfill \\
  \forall \left[ {\left( {q{'_i},{a_k},q{'_j},b{'_k},d'} \right),{F_1}\left( {{\mu ^t}(q{'_i}),\delta (q{'_i},{a_k},q{'_j},b{'_k},d')} \right)} \right] \in \Delta _{Act}^t({a_k}), \hfill \\
  \sum {\left\lceil {{F_1}\left( {{\mu ^t}(q{'_i}),\delta (q{'_i},{a_k},q{'_j},b{'_k},d')} \right)} \right\rceil }  < \sum {\left\lceil {{F_1}\left( {{\mu ^t}({q_i}),\delta ({q_i},{a_k},{q_j},{b_k},d)} \right)} \right\rceil }  \hfill \\ 
\end{gathered}  \right\}
\label{F4Formula3}
\end{equation}
\end{strip}

\end{itemize}

\section{\textbf{Comprehensive Fuzzy Turing Machine}}

Based on the discussions of the issues of the conventional FTM and the presented solutions for each issue, it is the time to define complete version of  CFTM:

\begin{definition}(Comprehensive Fuzzy Turing Machine, CFTM)\\
A Comprehensive Fuzzy Turing Machine is a single tape 4-tuple fuzzy Turing machine $\textbf{M}$ denoted as $\textbf{M}=\left( \textbf{T},F, \tilde \delta, \mu \right) $ which are defined as follows:
\begin{itemize}
\item  $\textbf{T}$ is the conventional fuzzy Turing machine which includes:
\begin{itemize}
\item $Q$ is the finite set of states.
\item $\Sigma$ is the finite set of tape symbols to be printed on the tape that has a leftmost cell, but it is unbounded to the right. 
\item $D$ is the set of possible head movement directions.
\item $I$ is the set of input symbols; $I \subset \Sigma$.
\item $\Delta_{CFTM}$ is is the next-move relation which is a subset of $ Q \times \Sigma \times Q \times \Sigma \times D$. For
each possible move of $\textbf{F}$ there is an element $\delta \in \Delta$ with $\delta = (q_1; a_1; q_2; a_2; d)$. That
is, if the current state is $q_1$ and the tape symbol scanned by the machine’s head is
$a_1$; $\textbf{F}$ will enter the new state $q_2$, the new tape symbol $a_2$ will rewrite the previous
symbol $a_1$, and the tape head will move in direction $d$. 
\item $B \in Q-I$ is the blank symbol. 
\item $\tilde R$ is the set of start states.
\item $Q_f$ is the set of final states. 
\end{itemize}
\item $\textbf{F}$ is the set of functions which includes:
\begin{itemize}
\item $F_1: [0,1] \times [0,1] \rightarrow [0,1]$ is the mapping function which is applied via $\tilde \delta$ to assign \textit{mv}s to the active states, thus called \textit{membership assignment function}. 
\item $\tilde \delta: (Q \times [0,1]) \times \Sigma \times Q \times \Sigma \times D   \xrightarrow{{{F_1}(\mu ,\delta )}} [0,1]$ is the augmented transition function. Please refer to section \ref{FTMMembAssign} for more details. 
\item $F_2:[0,1]^* \rightarrow [0,1]$ is a multi-membership  resolution function which resolves multi-membership active states and assigns a single \textit{mv} to them, thus called multi-membership resolution function. 
\item $F_3: \Delta^t_{Act}(a_k)  \rightarrow \Sigma$ is multi-symbol resolution function of the tape symbols to be printed on the tape during the FTM computations. $F_3$ assigns a single selected symbol to be printed on the tape at time $t$. $ \Delta_{Act}$ is the  set of current active transitions. 
\item $F_4: \Delta^t_{Act}(a_k)  \rightarrow D$ is multi-direction resolution function of the head movements during the FTM computations. $F_4$ determines a single direction for the head of the FTM tape to move at time $t$.
\end{itemize}
\item $\delta: \Delta_{CFTM} \rightarrow [0,1]$ is a function that  assigns transition weight in $[0,1]$ to each transition.$\square$
\item $\mu$ is the array of states membership values. 
\end{itemize}
\end{definition}

Conventionally, each FTM comes with its view on the concept of instantaneous description of the machine. In the definition below, the ID for our novel machine is presented:
\begin{definition}
Instantaneous Description (ID) of  Comprehensive Fuzzy Turing Machine (CFTM) $\textbf{M}$ working on the string $w$ at time $t \geqslant 0$ represented as $Q_t$ is defined as a unique description of the machine's tape, a vector of membership values of all CFTM states, and the position of the machine's head after performing the $t$-th move on the input $w$. 

\end{definition}

\begin{algorithm}[t]
 \KwData{
 
  The FTM Information\\
 The Tape Information
}
 \KwResult{The Membership Vales of All States after Entering the Input String.}
 Initialization\;
 \While{Not reached to the End of the Tape}{
  $Input Symbol$ = Read Tape Symbol\;
\For{All Transitions in FTM}{
   \If{Transition = Active}{
   Calculate F1 of that Transition\;
   Add $[Transition,F_1]$ pair to $\Delta^t_{Act}(Input Symbol)$ set\;
   }}
  \For{All States in Automata of FTM}{
  
   \uIf{Single membership value exists for a state}
   {
   Determine the membership value of the successor state at time step $(t+1)$\;}
  \uElseIf{Multi-Membership}
  {
   Do MultiMembership Resolution via calculation of $F_2$ for  the successor state at time step $(t+1)$\;
   }
  \textbf{end}
  }
    \For{All members of  $\Delta^t_{Act}(Input Symbol)$ set}{
  		Calculate the $F_3$ for Multi-Symbol Resolution to determine the Next Symbol to be written on the tape\;
		Calculate the $F_4$ for Multi-Direction Resolution to determine the Next Direction\;
  }
 }
 \caption{Pseudocode for  CFTM Calculations}
 \label{CFTMAlgorithm}
\end{algorithm} 

\begin{definition}\label{AccepDefLAb} (Acceptance) \\
A string is said to be accepted by a CFTM if and only if the membership value of at least one final state is not zero after the machine halts. Otherwise, the string is a member of a language which is  not supported by the Turing machine. The membership value of the final state is considered as ``Truth Degree'' or ``Acceptance Degree'' of that string processed by the CFTM.

\end{definition}
\textbf{Notice: } In cases that there are more than one final state with nonzero membership values, the multi-membership resolution is required again to determine the acceptance degree. The same conditions and definitions for $F_2$ is required or one might simply use the same $F_2$ in CFTM.

\begin{exmp}

An explanatory example of the computations of the  CFTM of Fig. \ref{FTMExampleNDterm_Lab} comes here. 
The CFTM  $\textbf{M}$ includes:\\
 $Q = \{ {q_0},{q_1},{q_2},{q_3},{q_4},{q_5}\}$,
 $\Sigma  = \{ 0,1\}$,
  $\Gamma  = \{ 0,1,B\}$,
   $q_0=$ start state,
    $\tilde R = \{ {q_2},{q_4}\}$. Also, suppose that the FTM starts with the state $q_0$ with membership value $1$.
 ${F_1} = {{(\mu  + \delta )} \mathord{\left/
 {\vphantom {{(\mu  + \delta )} 2}} \right.
 \kern-\nulldelimiterspace} 2}$,
  $F_2=\sqrt[n]{{{{\tilde \delta }_1} \times ... \times {{\tilde \delta }_n}}}$, and 
    $F_3$ and $F_4$ are symbol and direction with maximum cardinalities, respectively.\\

\begin{figure*}[th]
	\centering
  \includegraphics[width=0.95\textwidth]{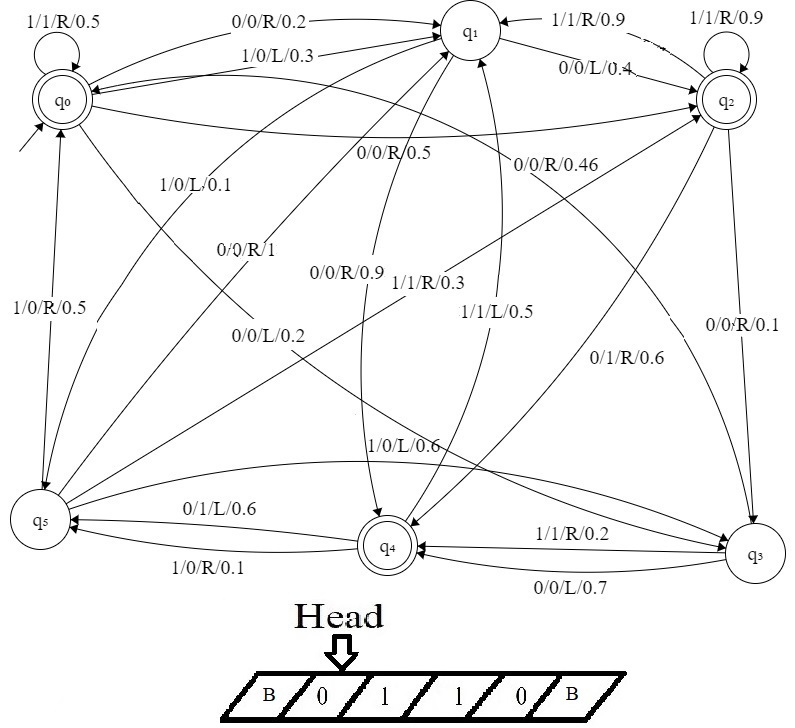}\\
  \caption{A Nondeterministic Fuzzy Turing Machine }
  \label{FTMExampleNDterm_Lab}
\end{figure*}

At time $t=0$ the ID of the CFTM, $Q_0$, is:\\
$\mu(states)=[1,0,0,0,0,0]$,
tape state: \begin{tabular}{|c||c||c|c|c|c|}
\hline 
\rule[-2ex]{0pt}{5.5ex}  B & 0 & 1 &1 & 0 &  B \\ 
\hline 
\end{tabular}\\
Head position is at cell $1$ -hypothetically the tape cells are numbered from 0.\\
The symbol read from the tape is ``$0$''.\\
\[\Delta _{Act}^0(0) = \left\{ \begin{gathered}
  \left[ {({q_0},0,{q_1},0,R),0.6} \right], \hfill \\
  \left[ {({q_0},0,{q_3},0,L),0.6} \right], \hfill \\
  \left[ {({q_0},0,{q_2},0,R),0.75} \right] \hfill \\ 
\end{gathered}  \right\}\]\\
Next head movement direction: $R$, and
the symbol written on the tape: $0$\\
No state requires multi-membership resolution.\\

At time $t=1$ the ID of the CFTM, $Q_1$, is:\\
$\mu(states)=[0.6,0.75,0.6,0,0,0]$,
tape state: \begin{tabular}{|c|c||c||c|c|c|}
\hline 
\rule[-2ex]{0pt}{5.5ex}  B & 0 & 1 &1 & 0 &  B \\ 
\hline 
\end{tabular} \\
Head position is at cell $2$, and the symbol read from the tape is ``$1$''.
$\begin{gathered}
  \Delta _{Act}^1(1) =  \hfill \\
  \left[ {({q_1},1,{q_5},0,R),0.55} \right],\left[ {({q_2},1,{q_2},1,R),0.825} \right], \hfill \\
  \left[ {({q_2},1,{q_1},1,R),0.825} \right],\left[ {({q_3},0,{q_0},0,R),0.53} \right], \hfill \\
  \left[ {({q_3},0,{q_4},0,L),0.65} \right] \hfill \\ 
\end{gathered} $
Next head movement direction: $R$, and the symbol written on the tape: $1$\\
No state requires multi-membership resolution.\\

At time $t=2$ the ID of the CFTM, $Q_2$, is:\\
$\mu(states)=[0.53,0.825,0.825,0,0.65,0.55]$,
tape state: \begin{tabular}{|c|c|c||c||c|c|}
\hline 
\rule[-2ex]{0pt}{5.5ex}  B & 0 & 1 &1 & 0 &  B \\ 
\hline 
\end{tabular} \\
Head position is at cell $3$, and the symbol read from the tape is ``$1$''.\\
\[\begin{gathered}
  \Delta _{Act}^2(1) =  \hfill \\
  \left\{ \begin{gathered}
  \left[ {({q_5},1,{q_2},1,R),0.425} \right],\left[ {({q_5},1,{q_3},0,L),0.575} \right], \hfill \\
  \left[ {({q_2},1,{q_2},1,R),0.862} \right],\left[ {({q_3},1,{q_4},1,R),0.425} \right], \hfill \\
  \left[ {({q_0},1,{q_0},1,R),0.515} \right],\left[ {({q_1},1,{q_5},0,L),0.462} \right], \hfill \\
  \left[ {({q_5},1,{q_0},0,R),0.775} \right],\left[ {({q_2},1,{q_1},1,R),0.862} \right], \hfill \\
  \left[ {({q_0},1,{q_1},0,L),0.415} \right] \hfill \\ 
\end{gathered}  \right\} \hfill \\ 
\end{gathered} \]

Next head movement direction: $R$, and the symbol written on the tape: $1$\\
There are some states that require multi-membership resolution:\\
For $q_0$, there are two membership value candidates: $\{0.515,0.775\}$ \\
$\sqrt {0.515 \times 0.775}  = 0.637$ \\
For $q_1$, there are two membership value candidates: $\{0.415,0.862\}$ \\
$\sqrt {0.415 \times 0.862}  = 0.598$ \\
For $q_2$, there are two membership value candidates: $\{0.425,0.862\}$ \\
$\sqrt {0.425 \times 0.862}  = 0.605$ \\

At time $t=3$ the ID of the CFTM, $Q_3$, is:\\
$\mu(states)=[0.637,0.598,0.605,0.575,0.425,0.462]$, tape state: \begin{tabular}{|c|c|c|c||c||c|}
\hline 
\rule[-2ex]{0pt}{5.5ex}  B & 0 & 1 &1 & 0 &  B \\ 
\hline 
\end{tabular} \\
Head position is at cell $4$ and the symbol read from the tape is ``$0$''.\\

\[\begin{gathered}
  \Delta _{Act}^3(0) =  \hfill \\
  \left\{ \begin{gathered}
  \left[ {({q_3},0,{q_0},0,R),0.517} \right],\left[ {({q_2},0,{q_4},1,R),0.602} \right], \hfill \\
  \left[ {({q_1},0,{q_2},0,L),0.499} \right],\left[ {({q_5},0,{q_1},0,R),0.731} \right], \hfill \\
  \left[ {({q_0},0,{q_2},0,R),0.565} \right],\left[ {({q_1},0,{q_2},0,L),0.499} \right], \hfill \\
  \left[ {({q_4},0,{q_5},1,L),0.512} \right],\left[ {({q_3},0,{q_4},0,L),0.637} \right], \hfill \\
  \left[ {({q_2},0,{q_3},0,R),0.352} \right],\left[ {({q_1},0,{q_4},0,R),0.749} \right], \hfill \\
  \left[ {({q_0},0,{q_1},0,R),0.415} \right],\left[ {({q_0},0,{q_3},0,L),0.415} \right] \hfill \\ 
\end{gathered}  \right\} \hfill \\ 
\end{gathered} \]

Next head movement direction: $R$, and the symbol written on the tape: $0$\\
There are some states that require multi-membership resolution:\\
For $q_1$, there are two membership value candidates: $\{0.731,0.415\}$ \\
$\sqrt {0.731 \times 0.415}  = 0.550$ \\
For $q_2$, there are three membership value candidates: $\{0.499,0.565,0.499\}$ \\
$\sqrt[3]{{0.499 \times 0.499 \times 0.565}}$ \\
For $q_3$, there are two membership value candidates: $\{0.352,0.415\}$ \\
$\sqrt {0.352 \times 0.415}  = 0.382$ \\
For $q_4$, there are three membership value candidates: $\{0.602,0.637,0.749\}$ \\
$\sqrt[3]{{0.602 \times 0.637 \times 0.749}}=0.659$ \\

For the next time step:\\
$\mu(states)=[0.517, 0.550,0.605,0.382,0.659, 0.512]$, and tape state for the next time step: \begin{tabular}{|c|c|c|c|c||c||}
\hline 
\rule[-2ex]{0pt}{5.5ex}  B & 0 & 1 &1 & 0 &  B \\ 
\hline 
\end{tabular}. \\

As we reached the end of string here, the CFTM enters the halt mode. It means the machine no longer works and the above configuration, $Q_3$ is actually the \textit{Final ID}. To determine whether the string ``$0110$'' is accepted, we refer to the  membership value of the final states, i.e. $q_0$, $q_2$, and $q_4$ in the final ID, $Q_3$.
The calculated membership values for these states at final ID are $\{0.517,0.605,0.659 \}$. To assign  a truth degree to the input string, again we face the multi-membership problem. Referring to the definition \ref{AccepDefLAb}, we might utilize the same definition for $F_2$ used throughout the calculations to resolve multi-membership problem for the acceptance degree. Therefore, the calculated  ``\textit{Acceptance Degree}'' of the string ``$0110$'' in CFTM   $\textbf{M}$ will be  $=\sqrt[3]{{0.517 \times 0.605 \times 0.659}}=0.590$.

\end{exmp}

Our suggested algorithm of CFTM computations  is as presented in Algorithm. \ref{CFTMAlgorithm}. Also, we made the source code for computing CFTM is available in Python  which can be found in \cite{NajmehThesis}.

\section{Conclusion}
In this paper, we instigated the conventional definition of FTM for their benefits and weaknesses. We noticed that the membership assignment is performed ID-based. In the light of General Fuzzy Automata (GFA) proposed by \cite{doostfatemeh_new_2005}, we developed a more complete definition for two problems already existed in fuzzy Turing machines which covers those vague aspects of the membership assignment and multi-membership resolution issue. we noticed that in FTMs, the membership assignment is not the only vague issue. Each active transition requires the machine to move its head in a specific direction and also mandates a predefined symbol to be written on the tape. Therefore, at each time step it is usually more than one symbol to be written on the tape and also more than one direction for the machine to move. 
Hence we defined two more functions to resolve the above mentioned issues, multi-direction     and multi-symbol resolution functions to decide on  a single direction and 
a single head movement based on the weight of the active transitions and the membership values of their predecessor states. It is easy to prove that each conventional fuzzy Turing machine can be modeled in the form of the novel Comprehensive Fuzzy Turing Machine (CFTM).

Lastly, using an example, some comparison on the volume of calculations on conventional FTM and the novel CFTM is performed. It is clear that the CFTM significantly reduces the amount of computations required for fuzzy Turing machine.


\bibliographystyle{plain}

\bibliography{JRNAL}

\end{document}